\documentclass[pdftex,twocolumn,epjc3]{svjour3}          

\RequirePackage[T1]{fontenc}

\smartqed  
\RequirePackage{graphicx}
\RequirePackage{mathptmx}      
\RequirePackage{flushend}
\RequirePackage[numbers,sort&compress]{natbib}
\RequirePackage[colorlinks,citecolor=blue,urlcolor=blue,linkcolor=blue]{hyperref}

\begin{document}

\title{Coupled quintessence and the impossibility of an interaction: a dynamical analysis study}


\author{Fabr\'izio F. Bernardi \thanksref{e1,addr1}
        \and
        Ricardo G. Landim\thanksref{e2,addr1} 
}

\thankstext{e1}{bernardiff@gmail.com}
\thankstext{e2}{rlandim@if.usp.br}

\institute{Instituto de F\'isica, Universidade de S\~ao Paulo\\
 Caixa Postal 66318,  05314-970 S\~ao Paulo, S\~ao Paulo, Brazil\label{addr1}
}

\date{Received: date / Accepted: date}

\maketitle

\begin{abstract}
We analyze the coupled quintessence in the light of the linear dynamical systems theory, with two different interactions: i) proportional to the energy density of the dark energy and ii) proportional  to the sum of the energy densities of the dark matter and dark energy. The results presented here enlarge the previous analyses in the literature, wherein the interaction has been only proportional to the energy density of the dark matter. In the first case it is possible to get the well-known sequence of cosmological eras. For the second interaction only the radiation and the dark energy era can be described by the fixed points. Therefore, from the point-of-view of the dynamical system theory, the interaction proportional  to the sum of the energy densities of the dark matter and dark energy does not describe the universe we live in.
  \end{abstract}

\section{Introduction}

Sixty eight percent of our universe \cite{Ade:2015xua} consists of a still mysterious component called ``dark energy'' (DE), which is believed to be responsible for the  present acceleration of the universe \cite{reiss1998, perlmutter1999}. In addition to ordinary matter, the remaining $27\%$ of the energy content of the universe is a form of matter that interacts in principle only gravitationally, known as dark matter (DM). Among a wide range of alternatives for the dark energy,  which includes the cosmological constant, scalar or vector fields \cite{ArmendarizPicon:2000dh,Padmanabhan:2002cp,Bagla:2002yn,Brax1999,Copeland2000,Landim:2015upa,ArmendarizPicon:2004pm,Koivisto:2008xf,Bamba:2008ja,Emelyanov:2011ze,Emelyanov:2011wn,Emelyanov:2011kn,Kouwn:2015cdw}, holographic dark energy \cite{Hsu:2004ri,Li:2004rb,Nojiri:2005pu,Pavon:2005yx,Wang:2005jx,Wang:2005pk,Wang:2005ph,Wang:2007ak,Landim:2015hqa,Li:2009bn,Li:2009zs,Li:2011sd,Wang:2016och}, metastable dark energy \cite{Stojkovic:2007dw,Landim:2016isc,Greenwood:2008qp,Abdalla:2012ug,Shafieloo:2016bpk}, modifications of gravity and different kinds of cosmological fluids \cite{copeland2006dynamics, Nojiri:2010wj, Bamba:2012cp, dvali2000, yin2005,Jamali:2016zww,Capozziello:2013bma}, the usage of a canonical scalar field, called ``quintessence'', is a viable candidate \cite{peebles1988,ratra1988,Frieman1992,Frieman1995,Caldwell:1997ii}.

In addition, the two components of the dark sector may interact with each other \cite{Wetterich:1994bg,Amendola:1999er,Farrar:2003uw,Guo:2004vg,Cai:2004dk,Guo:2004xx,Bi:2004ns,Gumjudpai:2005ry,yin2005,Wang:2005jx,Wang:2005pk,Wang:2005ph,Wang:2007ak,micheletti2009,Costa:2014pba,Shahalam:2015sja,Nunes:2016dlj,Sola:2016ecz} (see \cite{Wang:2016lxa} for a review) and the interaction can eventually alleviate the coincidence problem \cite{Zimdahl:2001ar,Chimento:2003iea}. 

When a scalar field is in the presence of a barotropic fluid (with equation of state $w_m=p_m/\rho_m$, where $p_m$ is the pressure and $\rho_m$ is the energy density of the fluid)  the relevant evolution equations can be converted   into an autonomous system. Such approach is a good tool to analyze asymptotic states of cosmological models and it has been done for uncoupled dark energy (quintessence, tachyon field, phantom field and vector dark energy, for instance \cite{copeland1998,ng2001,Copeland:2004hq,Zhai2005,DeSantiago:2012nk,Azreg-Ainou:2013jxa,Landim:2016dxh,Alho:2015ila}) and coupled dark energy \cite{Amendola:1999er,Gumjudpai:2005ry,TsujikawaGeneral,amendola2006challenges,ChenPhantom,Mahata:2015lja,Landim:2015poa,Landim:2015uda}.  The coupling assumed for the quintessence field has been proportional to the energy density of the dark matter $\rho_m$. However, there are other possibilities as for instance the coupling proportional to the energy density of the dark energy $\rho_\phi$ or the sum of the two energy densities $\rho_m+\rho_\phi$. Similar kernels have been widely studied in the literature \cite{Abdalla:2007rd,He:2008tn,He:2008si,Valiviita:2008iv,Abdalla:2009mt,Gavela:2009cy,He:2010im,Marcondes:2016reb}. In particular, the dark energy evolution at high redshifts measured by the BOSS-SDSS Collaboration \cite{Delubac:2014aqe} shows a deviation from the cosmological constant which can be explained assuming interacting dark energy models \cite{Abdalla:2014cla}. 

A dynamical analysis remained to be done for these two kernels. In this paper we use  the linear dynamical systems theory to investigate the critical points that come from the evolution equations for the quintessence, assuming the interaction between DE and DM proportional to i) $\rho_\phi $ and ii) $\rho_\phi+\rho_m$. We found that in the case i) there are fixed points that can describe the sequence of three cosmological eras. In the second case either radiation era or dark energy era can be described by fixed points, but the matter-dominated universe is absent.

The remainder of this paper is structured as follows. In Sect. \ref{de} we present the basics of the interacting dark energy and the dynamical analysis theory. The quintessence dynamics is presented in Sect. \ref{quint} and the dynamical system theory is used to study the coupled quintessence in Sect. \ref{AS}, wherein the critical points are shown. We summarize our results in Sect. \ref{conclu}. We use Planck units ($\hbar=c =M_{pl}=1$) throughout the text.

\section{Interacting dark energy and dynamical analysis}\label{de}

We  consider that dark energy is described by a scalar field with energy density $\rho_\phi$ and pressure $p_\phi$, and with an equation of state  given by $w_\phi=p_\phi/\rho_\phi$. We assume that the scalar field is coupled with dark matter, in such a way that total energy-momentum tensor is still conserved. In the flat  Friedmann--Lema\^itre--Robertson--Walker (FLRW) background with a scale factor $a$, the continuity equations for both components and for radiation are

\begin{equation}\label{contide}
\dot{\rho_\phi}+3H(\rho_\phi+p_\phi)=-\mathcal{Q},
\end{equation}

\begin{equation}\label{contimatter}
\dot{\rho_m}+3H\rho_m=\mathcal{Q},
\end{equation}

\begin{equation}\label{contirad}
\dot{\rho_r}+4H\rho_r=0,
\end{equation}

\noindent respectively, where $H=\dot{a}/a$ is the Hubble rate,  $\mathcal{Q}$ is the coupling and the dot is a derivative with respect to the cosmic time $t$. The indices $m$ and $r$ stand for matter and radiation, respectively.\footnote{We could be more economic if we had written the matter and radiation equations in a joint form, as a general barotropic fluid with equation of state $w_{b}$. The results would be, of course, unchanged.} The case of $\mathcal{Q}>0$ corresponds to dark energy transformation into dark matter, while $\mathcal{Q}<0$ is the transformation in the opposite direction. In principle, the coupling  can depend on several variables $\mathcal{Q}=\mathcal{Q}(\rho_m,\rho_\phi, \dot{\phi},H,t,\dots)$, so we assume for the quintessence the coupling is i) $\mathcal{Q}=Q \rho_\phi\dot{\phi}$ and ii) $\mathcal{Q}=Q (\rho_\phi+\rho_m)\dot{\phi}$, where $Q$ is a positive constant. The case with negative $Q$ is the same as the case with $Q>0$ but with negative fixed point $x$, described in the next section by $\frac{\dot{\phi}}{\sqrt{6}H}$ (\ref{eq:dimensionlessXYS}).

To deal with the dynamics of the system, we will define dimensionless variables. The new variables are going to characterize a system of differential equations in the form

\begin{equation}
X'=f[X],
\end{equation}

\noindent where $X$ is a column vector of dimensionless variables and the prime is the derivative  with respect to $ \log a$, where we set the present scale factor $a_0$ to be one. The critical points $X_c$ are those ones that satisfy $X'=0$. In order to study stability of the fixed points we consider linear perturbations $U$ around them, thus $X=X_c+U$. At the critical point the perturbations $U$ satisfy the following equation

\begin{equation}
U'=\mathcal{J}U,
\end{equation}

\noindent where $\mathcal{J}$ is the Jacobian matrix. The stability around the fixed points depends on the nature of the eigenvalues ($\mu$) of $\mathcal{J}$, in such a way that they are stable points if they all have negative values, unstable points if they all have positive values and saddle points if at least one eigenvalue has positive (or negative) value, while the other ones have opposite sign.  In addition, if any eigenvalue is a complex number, the fixed point can be stable (Re $\mu<0$) or unstable (Re $\mu>0$) spiral, due to the oscillatory behavior of its imaginary part.

\section{Quintessence dynamics}\label{quint}

The  scalar field $\phi$ is described by the  Lagrangian

\begin{equation}\label{scalar}
 \mathcal{L}=-\sqrt{-g}\left(\frac{1}{2}\partial^\mu\phi\partial_\mu\phi+V(\phi)\right),
\end{equation} 

\noindent where $V(\phi)$ is the scalar potential  given by $V(\phi)=V_0e^{-\lambda \phi}$ and $V_0$ and $\lambda$ are constants. This choice is motivated by the autonomous system, as we shall see soon. For a homogeneous field $\phi\equiv\phi(t)$  in an expanding universe with FLRW metric with scale factor $a\equiv a(t)$, the equation of motion is

\begin{equation}\label{eqmotionscalar1}
 \dot{\phi}(\ddot{\phi}+3 H\dot{\phi}+V'(\phi))=0,
\end{equation}

\noindent where the prime denotes derivative with respect to $\phi$.  

The interaction between the quintessence field with DM enters in the right-hand side of Eq. (\ref{eqmotionscalar1}).

In the presence of matter and radiation, the Friedmann equations for the scalar field are

\begin{equation}\label{eq:1stFEmatterS}
  H^2=\frac{1}{3}\left(\frac{\dot{\phi}^2}{2}+V(\phi)+ \rho_m+\rho_r\right),
\end{equation}

\begin{equation}\label{eq:2ndFEmatterS}
  \dot{H}=-\frac{1}{2}\left(\dot{\phi}^2+\rho_m+\frac{4}{3}\rho_r\right),
\end{equation}

\noindent and the equation of state becomes

\begin{equation}\label{eqstateS}
 w_\phi=\frac{p_\phi}{\rho_\phi}=\frac{\dot{\phi}^2-2 V(\phi)}{\dot{\phi}^2+2 V(\phi)}.
\end{equation}

We are now ready to proceed to the dynamical analysis of the system.

\section{Autonomous system} \label{AS}

The dimensionless variables are defined as

\begin{eqnarray}\label{eq:dimensionlessXYS}
 x\equiv  &\frac{\dot{\phi}}{\sqrt{6}H}, \quad y\equiv \frac{\sqrt{V(\phi)}}{\sqrt{3}H},
  \quad z\equiv \frac{\sqrt{\rho_r}}{\sqrt{3}H}, \nonumber\\
	&\quad \lambda\equiv -\frac{V'}{V}, \quad \Gamma\equiv \frac{VV''}{V'^2}.
\end{eqnarray}

The dark energy density parameter is written in terms of these new variables as

\begin{equation}\label{eq:densityparameterXYS}
 \Omega_\phi \equiv \frac{\rho_\phi}{3H^2} = x^2+y^2,
 \end{equation}

\noindent so that Eq. (\ref{eq:1stFEmatterS}) can be written as 

\begin{equation}\label{eq:SomaOmegasS}
\Omega_\phi+\Omega_m+\Omega_r=1,
\end{equation}

\noindent where the matter and radiation density parameter are defined by $\Omega_i=\rho_i/(3H^2)$, with $i=m,r$. From Eqs. (\ref{eq:densityparameterXYS}) and (\ref{eq:SomaOmegasS}) we have that $x$ and $y$ are restricted in the phase plane by the relation

\begin{equation}\label{restrictionS}
0\leq x^2+y^2\leq 1,
 \end{equation}
 
\noindent due to $0\leq \Omega_\phi\leq 1$. The equation of state $w_\phi$  becomes

\begin{equation}\label{eq:equationStateXYS}
 w_\phi =\frac{x^2-y^2}{x^2+y^2},
\end{equation}

\noindent  and the total effective equation of state is

\begin{equation}\label{eq:weffS}
 w_{eff} = \frac{p_\phi+p_r}{\rho_\phi+\rho_m+\rho_r}=x^2-y^2+\frac{z^2}{3},
\end{equation}

\noindent with an accelerated expansion for  $w_{eff} < -1/3$.

\subsection{Interaction $Q\rho_\phi$}

The dynamical system for the variables  $x$,  $y$, $z$  and $\lambda$ with the interaction proportional to $\rho_\phi$ is

\begin{eqnarray}\label{dynsystemS}\label{eq:dx1/dnS}
\frac{dx}{dN}&=-3x+\frac{\sqrt{6}}{2}y^2\lambda-\frac{\sqrt{6}}{2} Q(x^2+y^2)
-xH^{-1}\frac{dH}{dN},\end{eqnarray}

\begin{equation}\label{eq:dy/dnS}
\frac{dy}{dN}=-\frac{\sqrt{6}}{2}x y\lambda-yH^{-1}\frac{dH}{dN},
\end{equation}

\begin{equation}\label{eq:dz/dnS}
\frac{dz}{dN}=-2z-zH^{-1}\frac{dH}{dN},
\end{equation}

\begin{equation}\label{eq:dlambda/dnS}
\frac{d\lambda}{dN}=-\sqrt{6}\lambda^2 x\left(\Gamma-1\right),
\end{equation}

\noindent where

\begin{equation}\label{}
H^{-1}\frac{dH}{dN}=-\frac{3}{2}\left(1+x^2-y^2+\frac{z^2}{3}\right).
\end{equation}

\subsubsection{Critical points}

 The fixed points of the system are obtained by setting $dx/dN=0$, $dy/dN=0$, $dz/dN=0$ and $d\lambda/dN=0$ in Eq. (\ref{dynsystemS})--(\ref{eq:dlambda/dnS}). When $\Gamma=1$, $\lambda$ is constant the potential is $V(\phi)=V_0e^{-\lambda \phi}$ \cite{copeland1998,ng2001}.\footnote{The equation for $\lambda$ is also equal zero when $x=0$ or $\lambda=0$, so that  $\lambda$ should not necessarily be constant, for the fixed points with this value of $x$. However, for the case of dynamical $\lambda$, the correspondent eigenvalue is equal zero, indicating that the  fixed points is not hyperbolic.} The fixed points are shown in Table \ref{criticalpointsS}. Notice that $y$ cannot be negative and recall that $\Omega_r=z^2$.

 \begin{table*}[t]\centering
\begin{tabular}{lllllll}
\hline\noalign{\smallskip}
Point   & $x$ & $y$  &$z$& $w_\phi$ & $\Omega_\phi$& $w_{eff}$ \\
\noalign{\smallskip}\hline\noalign{\smallskip}
(a)  & 	$\frac{2\sqrt{6}}{3\lambda}$   &$ \frac{2 \sqrt{2 Q + \lambda}}{\sqrt{3\lambda^2(\lambda-Q)}} $ & $\sqrt{1-\frac{4(\lambda-4Q)}{\lambda^2(\lambda-Q)}}$&$\frac{1}{3}\left(1-\frac{4Q}{\lambda}\right)$& $\frac{4}{\lambda(\lambda-Q)}$ &$\frac{1}{3}$ \\
		(b) &0 & 0  & $1$   &$-$& 0&$\frac{1}{3}$  \\
				 (c) & $-\frac{\sqrt{6}}{3Q}$ &0   &$\sqrt{1-\frac{2}{Q^2}}$&$1$&$\frac{2}{3Q^2}$&$\frac{1}{3}$ \\ 
		 (d) & $0$ & 0 &0 &$-$ &$0$&0\\
		 
    (e) & $\frac{Q\pm\sqrt{Q^2+6}}{\sqrt{6}}$&$0$   &0& $1$& $x_e^2$&  $x_e^2$\\ 
 
         (f)& $x_f$&$\sqrt{1+x_f^2-\sqrt{2}x_f \lambda}$   &0&$\frac{-3 + \sqrt{6} x_f \lambda}{3 + 6 x_f^2 - \sqrt{6} x_f \lambda}$&$1 + 2 x_f^2-\sqrt{\frac{2}{3}} x_f \lambda$&$-1 + \sqrt{\frac{2}{3}} x_f \lambda$ \\

 \noalign{\smallskip}\hline
\end{tabular}
\caption{\label{criticalpointsS} Critical points ($x$, $y$ and $z$) of the Eqs. (\ref{dynsystemS})--(\ref{eq:dz/dnS}) for the quintessence  field with interaction $Q\rho_\phi$. The table shows the correspondent equation of state for the dark energy (\ref{eq:equationStateXYS}), the effective equation of state (\ref{eq:weffS}) and the density parameter for dark energy (\ref{eq:densityparameterXYS}). }
\end{table*}

The point $x_f$ is

\begin{eqnarray}
x_f&=\frac{-3 + Q \lambda - \lambda^2 + \sqrt{
 Q^2 (\lambda^2-12) - 
  2 Q \lambda ( \lambda^2-9) + (\lambda^2-3 )^2}}{2 \sqrt{6} (Q - \lambda)}.
\label{eq:xf}
\end{eqnarray}

The eigenvalues of the Jacobian matrix were found for each fixed point in Table \ref{criticalpointsS}.  The results are shown in Table \ref{stabilityS}, where the eigenvalues $\mu_{f\pm}$ are 

\begin{eqnarray}
\mu_{f\pm}&=\frac{1}{4}\{-12 + \sqrt{6} x_f (-2 Q + 3 \lambda)\nonumber \\& \pm  [
   48 (Q - \lambda) \lambda + 96 \sqrt{6} x_f^3 (-Q + \lambda) \nonumber\\& +
    6 x_f^2 (-24 + 4 Q^2 + 28 Q \lambda - 31 \lambda^2) \nonumber\\& + 
    8 \sqrt{6}
      x_f (9 \lambda + 2 \lambda^3 - 2 Q (3 + \lambda^2))]^{1/2}\}
\label{eq:mu}
\end{eqnarray}

\begin{table*}
\centering
\begin{tabular}{lllll}
\hline\noalign{\smallskip}
  Point & $\mu_1$ & $\mu_2$ & $\mu_3$& Stability \\
\noalign{\smallskip}\hline\noalign{\smallskip}
  (a)        & & see the main text& &saddle  \\
  (b)    & $1$&$3-\sqrt{6}Q$&$3-\frac{\sqrt{6}}{2}\lambda$& saddle or unstable\\
  (c)  & $1-\frac{\sqrt{2}}{Q}$&$1+\frac{\sqrt{2}}{Q}$  &$2-\frac{\lambda}{Q}$&saddle or unstable\\ 
  (d)   & $-\frac{3}{2}$&$\frac{3}{2}$   &   $-\frac{1}{2}$ &saddle     \\ 
   (e)   & $\frac{1}{2}\left(2+Q\left(Q\pm\sqrt{6+Q^2}\right)\right)$&$\frac{1}{2}\left(6+Q\left(Q\pm \sqrt{6+Q^2}\right)\right)$   &  $\frac{1}{2}\left(6+\left(Q-\lambda\right)\left(Q\pm\sqrt{6+Q^2}\right)\right)$ &saddle   or unstable \\
   (f)  & $-2 + \sqrt{\frac{3}{2}} x_f \lambda, $&  $\mu_{f+}$& $\mu_{f-}$& stable \\
     
     \noalign{\smallskip}\hline
\end{tabular}
\caption{\label{stabilityS} Eigenvalues and stability of the fixed points for the quintessence field with interaction $Q\rho_\phi$. }
\end{table*}

The fixed point (a) describes a radiation-dominated universe and in order to the fixed points be real and $\Omega_\phi$ satisfy the  nucleosynthesis bound $\Omega_\phi^{BBN}<0.045$ \cite{bean2001} we should have $\lambda> \frac{20\sqrt{2}}{3}$ and $Q\leq\frac{9\lambda^2-800}{9\lambda}$. The eigenvalues were found numerically. For $\lambda=10$ and the the upper limit for the interaction ($Q=10/9$) we get the eigenvalues $\mu_1=-0.7+2.1 i$, $\mu_2=-0.7-2.1 i$ and $\mu_3=1$, so this critical point is a saddle point. Similar results are found for other values of $\lambda$ and $Q$.

Both points (b) and (c) also describe the radiation era and are unstable or saddle points. The eigenvalues $\mu_2$ and $\mu_3$ of the point (b) can be either positive or negative, depending on the values of $\lambda$ and $Q$. On the other hand, the first eigenvalue $\mu_1$ is always positive. The same happens with the eigenvalues of the point (c) and in this case the interaction must be $Q>\sqrt{2}$ for the fixed points be real and $Q>20/(3\sqrt{3})$ for  $\Omega_\phi^{BBN}<0.045$.

The matter-dominated universe is described by the saddle point (d) and also by the point (e), provided that $Q$ is sufficiently large for the latter case. Whatever the value of the interaction all eigenvalues of the point (e) cannot be simultaneously negative.

\begin{figure}
\centering
	\includegraphics[width=\columnwidth]{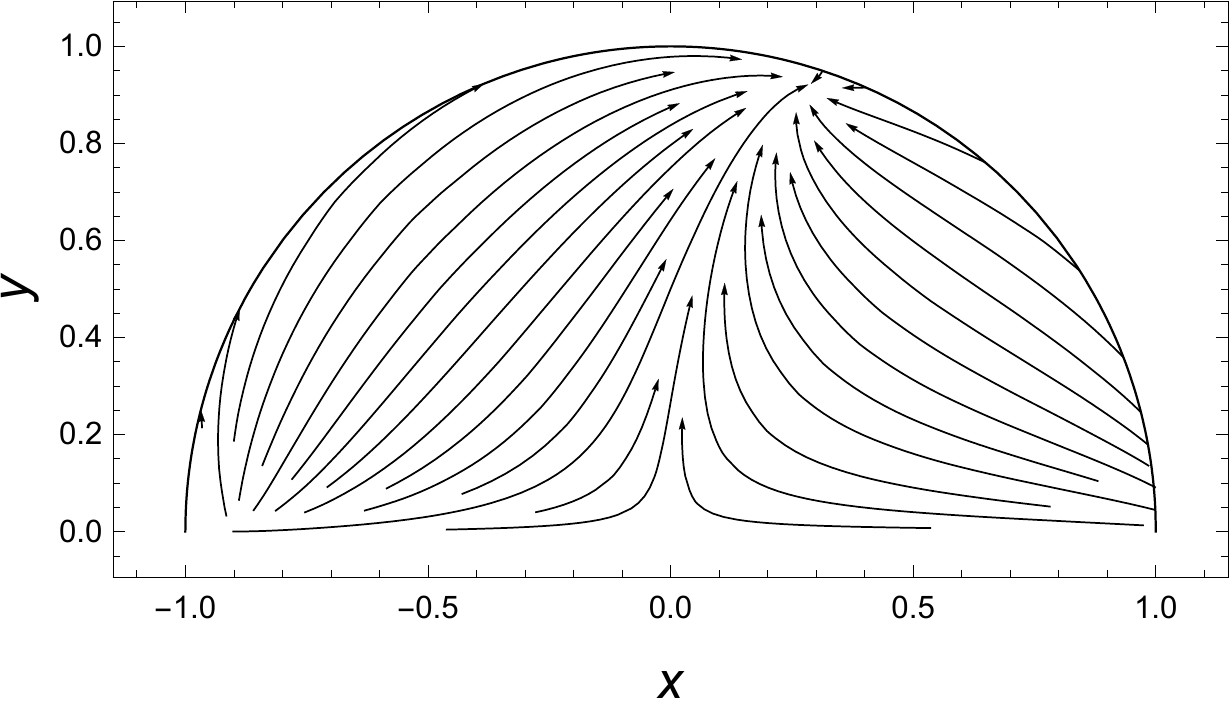}
	\caption{Phase plane for the fixed point (f) with $\lambda=1$ and $Q=1/4$.}
		\label{fig:PhasePlane}
\end{figure}

The last fixed point (f) is an attractor and  it describes the dark-energy dominated universe if either $\lambda \leq \sqrt{2}$ and  $Q < \lambda$ or $\lambda > \sqrt{2}$ and $\frac{-2 \lambda + \lambda^3}{  4 + \lambda^2}\leq Q < \lambda$. The real part of the eigenvalues are negative for these range of values, thus the fixed point is stable or stable spiral. Its behavior is illustrated in Fig. \ref{fig:PhasePlane}, where we plot the phase plane with $\lambda=1$ and $Q=1/4$.

\begin{figure}
\centering
	\includegraphics[scale=0.55]{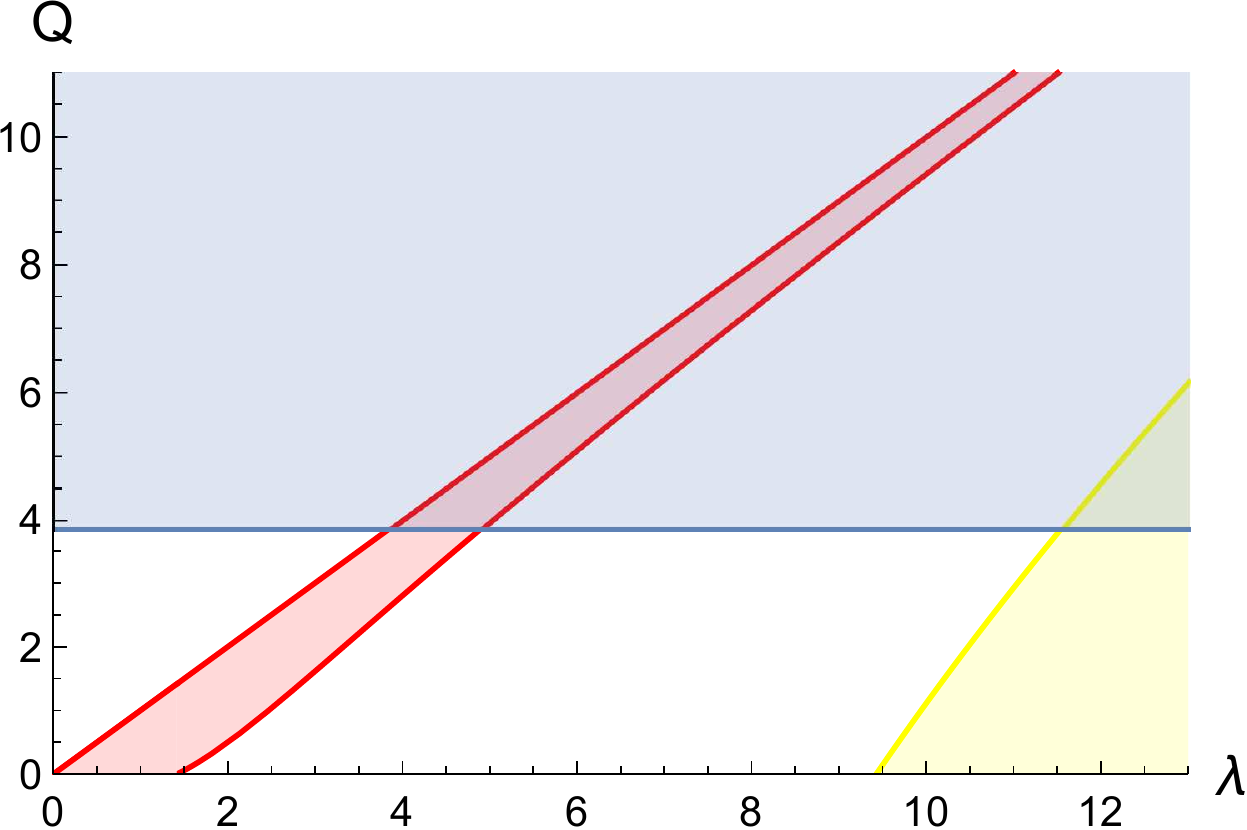}
	\caption{Allowed regions of $Q$, for the fixed points (a) (yellow), (c) (blue) and (f) (red). }
		\label{allowedreg}
\end{figure}

The allowed values of $\lambda$ and $Q$, for the fixed points (a), (c) and (f), are shown in Fig. \ref{allowedreg}. From the figure we see that the fixed points (a) and (f) do not have common regions. 

Therefore, the sequence of cosmological eras (radiation $\rightarrow$ matter $\rightarrow$ dark energy) is reached considering the transition: (b) or (c) $\rightarrow$ (d) or (e) $\rightarrow$ (f).

 
\subsection{Interaction $Q(\rho_\phi+\rho_m)$}

The dynamical system for the variables  $x$,  $y$, $z$  and $\lambda$ with the interaction proportional to $\rho_\phi+\rho_m$ is

\begin{eqnarray}\label{dynsystemS2}\label{eq:dx1/dnS2}
\frac{dx}{dN}&=-3x+\frac{\sqrt{6}}{2}y^2\lambda-\frac{\sqrt{6}}{2} Q(1-z^2)
-xH^{-1}\frac{dH}{dN},\end{eqnarray}

\begin{equation}\label{eq:dy/dnS2}
\frac{dy}{dN}=-\frac{\sqrt{6}}{2}x y\lambda-yH^{-1}\frac{dH}{dN},
\end{equation}

\begin{equation}\label{eq:dz/dnS2}
\frac{dz}{dN}=-2z-zH^{-1}\frac{dH}{dN},
\end{equation}

\begin{equation}\label{eq:dlambda/dnS2}
\frac{d\lambda}{dN}=-\sqrt{6}\lambda^2 x\left(\Gamma-1\right),
\end{equation}

\noindent where

\begin{equation}\label{}
H^{-1}\frac{dH}{dN}=-\frac{3}{2}\left(1+x^2-y^2+\frac{z^2}{3}\right).
\end{equation}

All equations above but the first one are identical to the previous case.

\subsubsection{Critical points}

 As before fixed points of the system are obtained by setting $dx/dN=0$, $dy/dN=0$, $dz/dN=0$ and $d\lambda/dN=0$ in Eq. (\ref{dynsystemS2})--(\ref{eq:dlambda/dnS2}).  The fixed points are shown in Table \ref{criticalpointsS2}, where

\begin{equation}
w_{\phi e}=\frac{\lambda^2\left(\lambda^2-3-\sqrt{12Q\lambda+(\lambda^2-3)^2}\right)}{3\left(\lambda(\lambda+2Q)+3-\sqrt{12Q\lambda+(\lambda^2-3)^2}\right)}.
\label{eq:wphie}
\end{equation}

 \begin{table*}\centering
\begin{tabular}{lllllll}
\hline\noalign{\smallskip}
Point   & $x$ & $y$  &$z$& $w_\phi$ & $\Omega_\phi$& $w_{eff}$ \\
\noalign{\smallskip}\hline\noalign{\smallskip}
(a)  & 	$\frac{2\sqrt{6}}{3\lambda}$   &$ \frac{2 \sqrt{6 Q + \lambda}}{\sqrt{3\lambda^2(\lambda+3Q)}} $ & $\sqrt{\frac{\lambda(\lambda+3Q)-4}{\lambda(\lambda+3Q)}}$&$\frac{\lambda}{3\lambda+12Q}$& $\frac{4((\lambda+4Q)}{\lambda^2(\lambda+3Q)}$ &$\frac{1}{3}$ \\
		(b) &0 & 0  & $1$   &$-$& 0&$\frac{1}{3}$  \\
				 (c) & $-\frac{\sqrt{6}}{9Q}$ &0   &$\sqrt{1-\frac{2}{9Q^2}}$&$1$&$\frac{2}{27Q^2}$&$\frac{1}{3}$ \\ 
		 (d) & $3x_d^3-3x_d-\sqrt{6}Q=0$ & 0 &0 &$1$ &$x_d^2$&$x_d^2$\\
		 
    (e) & $\frac{3+\lambda^2\pm\sqrt{12Q\lambda+(\lambda^2-3)^2}}{2\sqrt{6}\lambda}$&$\sqrt{x_e^2-(\sqrt{6}/3)x_e\lambda+1}$   &0& $w_{\phi e}$&  $\frac{\lambda^2+2Q\lambda+3-\sqrt{12Q\lambda+(\lambda^2-3)^2}}{2\lambda^2}$&  $\frac{\left(\lambda^2-3-\sqrt{12Q\lambda+(\lambda^2-3)^2}\right)}{6}$\\

 \noalign{\smallskip}\hline
\end{tabular}
\caption{\label{criticalpointsS2} Critical points ($x$, $y$ and $z$) of the Eqs. (\ref{dynsystemS2})--(\ref{eq:dz/dnS2}) for the quintessence  field with interaction $Q(\rho_\phi+\rho_m)$. The table shows the correspondent equation of state for the dark energy (\ref{eq:equationStateXYS}), the effective equation of state (\ref{eq:weffS}) and the density parameter for dark energy (\ref{eq:densityparameterXYS}). }
\end{table*}

The eigenvalues of the Jacobian matrix were found for each fixed point of the Table \ref{criticalpointsS2}.  The results are shown in Table \ref{stabilityS2}, where

\begin{equation}
\mu_{e1}=\frac{1}{4}\left(\lambda^2-5-\sqrt{\mu}\right),
\label{eq:3333}
\end{equation}

\begin{eqnarray}
\mu_{e2,e3}&=\frac{1}{8\lambda^2}(3\lambda^4+3\lambda^2(5-\sqrt{\mu})
\\\nonumber &\pm\sqrt{2}(\lambda^2(-72(-3+\sqrt{\mu}))\\\nonumber &-6Q\lambda(7\lambda^2-48+8\sqrt{\mu})\\\nonumber &+\lambda^2(\lambda^4-63-(3+\lambda^2)\sqrt{\mu}))^{1/2})
\label{eq:333333}
\end{eqnarray}

\noindent and

\begin{equation}
\mu=12Q\lambda+(\lambda^2-3)^2.
\label{eq:}
\end{equation}

\begin{table*}
\centering
\begin{tabular}{lllll}
\hline\noalign{\smallskip}
  Point & $\mu_1$ & $\mu_2$ & $\mu_3$& Stability \\
\noalign{\smallskip}\hline\noalign{\smallskip}
  (a)        & & see the main text& &saddle  \\
  (b)    & $2$&$-1$&$1$& saddle\\
  (c)  & $-\sqrt{\frac{2}{9Q^2}-3}$&$\sqrt{\frac{2}{9Q^2}-3}$  &$2+\frac{\lambda}{3Q}$&saddle\\ 
  (d)   & &see the main text &   &saddle  or unstable   \\ 
  (e)   &$\mu_{e1}$ &$\mu_{e2}$ & $\mu_{e3}$  &stable   \\ 
   
     \noalign{\smallskip}\hline
\end{tabular}
\caption{\label{stabilityS2} Eigenvalues and stability of the fixed points for the quintessence field with $Q(\rho_\phi+\rho_m)$. }
\end{table*}

The point (a) describes a radiation-dominated universe and in order to the fixed points be real and $\Omega_\phi$ satisfy the  nucleosynthesis bound $\Omega_\phi^{BBN}<0.045$ \cite{bean2001} we should have $ \frac{20\sqrt{2}}{3}<\lambda<\frac{40\sqrt{6}}{9}$ and $Q\leq\frac{9\lambda^3-800\lambda}{27\lambda^2-3200}$ or $\lambda\geq\frac{40\sqrt{6}}{9}$ for any value of positive $Q$. The eigenvalues were found numerically and similarly to the case of the previous section, the fixed point is a saddle point for the allowed values of $\lambda$.

The radiation era is also described by the points (b) and (c). They are saddle points and for (c) the interaction must be $Q\geq \frac{20\sqrt{3}}{27}$ in order not to conflict the nucleosynthesis bound.

The matter-dominated universe can be described by the point (d) but only if the interaction is zero, which in turn is known in the literature \cite{copeland2006dynamics}. 

The fixed point (e) can describe the current stage of accelerated expansion of the universe for some values of $Q$ and $\lambda$. The critical points are real with $0\leq\Omega_\phi\leq 1$ and $w_{eff}<-1/3$ for $0<\lambda\leq \sqrt{2}$ and $0<Q\leq\lambda$ or for $\lambda>\sqrt{2}$ and $\frac{\lambda^2-2}{3\lambda}<Q\leq\lambda$. For these ranges of $\lambda$ and $Q$ the real part of the eigenvalues are negative, so the point is stable or stable spiral. The attractor point has $\Omega_\phi=1$ and $w_\phi=w_{eff}=-1$ for $Q=\lambda$. 

Therefore, both radiation and dark-energy-dominated universe can be described by the fixed points, however, none of them represent 	the matter era.

 \section{Conclusions}\label{conclu}
 
In the light of the linear dynamical systems theory we have studied coupled quintessence with dark matter with two different interactions: i) proportional to the energy density of the dark energy $\rho_\phi$ and ii) proportional  to the sum of the two energy densities $\rho_m+\rho_\phi$. The results presented here enlarge the previous analysis in the literature, wherein the interaction has been only proportional to the energy density of the dark matter. In the case i) the transition of cosmological eras is fully achieved with a suitable sequence of fixed points.  In the second case either radiation era or dark energy era can be described by the fixed points, but  not the matter-dominated universe. Therefore,  the second  interaction does not provide the  cosmological sequence: radiation $\rightarrow$ matter $\rightarrow$ dark energy. This is not the first time that an interaction proportional to the sum of the energy densities leads to cosmological disasters. A phenomenological model with that coupling suffers early-time instability for $w_d>-1$, as shown in \cite{Valiviita:2008iv,He:2008si}. Further analysis  for high redshifts and different coupling are summarized in \cite{Wang:2016lxa}.

\begin{acknowledgements}
 We thank Elcio Abdalla for comments. This work is supported by CAPES and CNPq. \end{acknowledgements}

\bibliographystyle{unsrt}
\bibliography{trab1}\end{document}